\documentclass[floatfix,preprintnumbers,superscriptaddress,showpacs,
               showkeys,prb,preprint]{revtex4}% twocolumn / preprint

\usepackage{graphicx}% Include figure files
\usepackage{dcolumn}% Align table columns on decimal point
\usepackage{amsmath, amsfonts, amssymb, bm}% bold math

\begin{document}

\bibliographystyle{apsrev} 
\preprint{A/FeHmd; submitted to Phys. Rev. B}

\title{Ab-initio molecular dynamics simulation of
hydrogen diffusion in $\alpha$-iron
} 

\author{J. Sanchez}
\author{J. Fullea}
\author{M.C. Andrade}

\affiliation{
Instituto de Ciencias de la Construccion Eduardo Torroja \\
}

\author{P.L. de Andres}

\affiliation{
Instituto de Ciencia de Materiales de Madrid  \\
Consejo Superior de Investigaciones Cientificas,
Madrid (SPAIN)
}

\date{\today}

\begin{abstract}
First-principles atomistic molecular dynamics simulation in the
micro-canonical and canonical ensembles has been used to 
study the diffusion of interstitial hydrogen in $\alpha$-iron. 
Hydrogen to Iron ratios between $\theta=\frac{1}{16}$ and $\frac{1}{2}$ 
have been considered by locating interstitial hydrogen atoms at random 
positions in a $2 \times 2 \times 2$ supercell. 
We find that the average optimum absorption site and the barrier for 
diffusion depend on the concentration of interestitials.
Iron Debye temperature decreases monotonically 
for increasing concentration of interstitial hydrogen, proving that iron-iron
interatomic potential is significantly weakened in the presence of a large 
number of diffusing hydrogen atoms. 
\end{abstract}


% 62.20.-x Mechanical properties of solids. 
% 62.20.mj Brittleness 
% 68.43.Bc ab initio calculation of adsorbates structure and reactions
% 68.43.Fg binding sites, geometry
% 66.30.J- 
% 66. nonelectronic transport properties of condensed matter
% 66.30.-h diffusion in solids 
% 66.30.J- diffusion of impurities  
% 68.43.Jk Diffusion of adsorbates, kinetics of coarsening and aggregation
% 71.15.-m Methods of electronic structure calculations
% 71.15.Mb Density functional theory, lda, gga
% 71.15.Nc Total energy and cohesive energy calculations


%Use showkeys class option if keyword display desired

\maketitle

%\section{\label{sec:intro} Introduction}
{\it Introduction.}
Diffusion of hydrogen atoms in metals make an impact on  a number of technologically
important properties of the host material and has attracted interest
since long ago.\cite{alefeld78IyII}
In particular, it is believed to play a crucial role in embrittlement
of high strength steels, of common use in buildings, bridges, 
etc.\cite{eliaz02,liang03, elices04b,scully04}
However, a fundamental understanding of different mechanisms linking embrittlement
and diffusion of hydrogen is still lacking owing to the complexity of the relevant
processes. 
Different authors have used Density Functional Theory (DFT) to study
preferred adsorption sites and diffusion barriers for hydrogen in
body-center cubic iron ($\alpha$-Fe), which constitutes the core of high-strength
steels and consequently is the simplest model for fundamental studies trying
to connect macroscopic and microscopic properties. 
\cite{norskov82,zheng89,louie98,hoffman99}
The main stream in the literature favors 
diffusion barriers of $\approx 0.1$ eV, and
adsorption of hydrogen
in the tetrahedral site which is the mot favorable one simply in terms
of available space.
Comprehensive work along these lines, including the effect of surfaces,
has been done by Jiang and Carter,
and and more recently Ramasubramaniam et al.\cite{carter04,carter08}
For large concentrations, however, we have found that occupancy of
the competing octahedral site, and reduced diffusion barriers, can 
happen.\cite{SanchezPRB08}
This is accompanied by a tetragonal distortion in the bcc unit cell and it is
related to internal stresses appearing upon an increasing hydrogen load
inside the system. 
From an experimental point of view, it has been observed that for
increasing external pressures of hydrogen the iron melting point is
lowered appreciably and the transition from $\alpha$ (bcc) to 
$\gamma$ (fcc) happens at lower temperature.\cite{fukai05}
Furthermore, neutron diffraction analysis on ferrite samples of
high-strenght steels have revealed a significant increase of 
Debye-Waller factors upon increasing loads of hydrogen.\cite{castellote07}
In this paper we investigate the effect of accumulating interstitial
atoms in a region, which is known to happen at least locally around
defects prior to the breaking of samples. 
We use ab-initio molecular dynamics because it presents several
advantages to study this problem. First, it allows us to study the many-body
interactions between simultaneously diffusing interstitial impurities. 
Second, temperature dependent simulations allow us to extract useful additional 
dynamical information. Third, for such a complex situation where the number of
possibilities to distribute a number of interstitial impurities among several candidate
sites becomes combinatorially large, a statistical approach where
the system follows its own internal dynamics to sample
relevant configurations is the more effective
approach.
Kinetic Monte-Carlo has 
also been recently applied to study the variation of barriers with different 
configurations and their associated stresses:\cite{carter08}
this is an alternative approach where good agreement with prior
ab-initio DFT calculations has been found,
proving the viability of statistical methods to study this kind of problem. 

{\it Methods.}
First-principles MD 
calculations have been performed with the 
CASTEP code. \cite{payne05,accelrys}
A $2 \times 2 \times 2$ periodic supercell is set up including sixteen Fe atoms and
several H atoms depending on the different 
concentrations being considered.
The Born-Oppenheimer approximation is used; ions are considered classical 
objects moving on the potential created by the electrons obeying the 
Schroedinger equation. Electronic wave-functions are expanded using a 
plane-waves basis set up to an energy cutoff of 375 eV and are sampled 
inside the Brillouin zone in a 
$4 \times 4 \times 4$ Monkhorst-Pack mesh.\cite{monk}
Electronic energies are converged up to $10^{-5}$ eV.
Ultra-soft pseudopotentials are used to describe Fe and 
H\cite{vanderbilt90} and the generalized gradients approximation for 
exchange and correlation due to Perdew, Burke and Ernzerhof is 
chosen\cite{PBE}. These approximation have been thoroughly checked 
before and it has been found that they reproduce correctly the main 
physical properties of $\alpha$-iron, including lattice constant, 
magnetization and bulk modulus.\cite{SanchezPRB08}
% referee point 3
A word of caution is in order here since we have used a k-point mesh less dense than the ones we have previously shown to be adequate to accurately reproduce different equilibrium properties of $\alpha$-iron at $T=0$ K.  Our choice is based in two different reasons. First of all, it is a practical one since molecular dynamics is a computer intensive task and a compromise between accuracy and time must be made. 
Secondly, we notice that computing different physical magnitudes require different accuracies, as can be deduced from Fig. 2 in ref.\cite{SanchezPRB08}; computing the equilibrium lattice parameter with a k-point mesh similar to the one used here incurs in an error of $0.4$\%, but to $\approx 300$\% for the bulk modulus. Our primary goal in this work is to compute total energies that can fluctuate in the current range of temperatures by
$\approx 0.05$ eV. We have checked that for the maximum density of interstitial impurities considered here the total energy changes at $T=0$ K by $0.047$ eV if the $4 \times 4 \times 4$ mesh is replaced by a  $8 \times 8 \times 8$ one. Therefore, the error is close to the random fluctuations intrinsic to the system and can be accepted. Furthermore, we notice that the error in the equilibrium lattice parameter by such an approximation is $\approx 0.01$ {\AA} (e.g. Fig. 2 in ref.\cite{SanchezPRB08}), which is of the same order of magnitude or lower than the root mean squared amplitude displacements of vibrating atoms at the temperatures used here. Therefore, we assume that the $4 \times 4 \times 4$ is both practical and accurate enough for the purpose of these simulations.
Finally, our current MD simulations predict 
a value for the Debye temperature of bcc iron ($\Theta=505$ K) 
in reasonable agreement with the reported experimental value ($420$ K).
% pt 3

To study the diffusion of several H atoms inside the unit cell we test 
in the micro-canonical ensemble the quality of the total energy 
conservation during a typical MD run. 
Simulations for 1 or 2 ps are performed with time steps of 0.5 or 1.0 fs 
showing that the total energy is conserved within a 0.01\% error 
(equilibration taking place during the first 100 to 200 fs have been 
taken out from averages). 
Keeping fixed the parameters defining the model, we switch to the 
canonical ensemble to reproduce conditions where the Fe and H atoms 
are in equilibrium with a thermal bath kept at a fixed temperature 
(Nose-Hover prescription has been used). 
All these simulations are performed keeping constant the volume of the
unit cell and the number of particles.
% referee point 4
% 0 179.64; 2 187.42; 4 191.77; 8 206.15
These boundary conditions are important to understand the physical model and the solutions obtained. In particular, it is relevant to discuss the meaning of keeping the volume of simulation fixed.  In a previous work we have investigated the volume deformation and atomic displacements necessary to find an equilibrium solution with
zero forces and stresses in the presence of interstitial hydrogen binding to either the octahedral or tetrahedral high symmetry sites.\cite{SanchezPRB08} Here we are interested, for the case of an overall low dilution concentration of impurities, in the effect of a high concentration of interstitial atoms inside a small region embedded in a matrix of iron that except for the large concentration of interstitial hydrogen in a small number of regions mostly keeps its original properties.  This is consistent with the experimental observation of
overall low dilution of interstitial H in $\alpha$-iron,  but possibly large concentration in particular  regions, maybe as a consequence of the existence of 
defects.\cite{alefeld78IyII,castellote07} Therefore, it is assumed here that the modification of the volume in the region of interest due to the internal  pressure created by the impurities is effectively controlled by the large amount of unperturbed bulk material surrounding the region where the interstitial hydrogen diffuses in a scale of time compatible with our simulations (ps).  Consequently, we fix the 
$2 \times 2 \times 2$ supercell volume used for the simulation to the one corresponding to pure $\alpha$-iron independently of the interstitial concentration inside. In the other limit a completely different model could be considered: an overall concentration distributed uniformly and large enough large enough so the volume should be adjusted for each density. For this case, the fractional variation of the supercell volume would be
4\%, 7\% and 15\$ for $n_{H}=2,4,8$, respectively (the restriction
$\alpha=\beta=\gamma=90^{\circ}$ has been used, but $a$, $b$, and $c$ have been allowed to change freely to minimize stresses).  While the (NVE) and (NVT) collectives are
appropriate for the first scenario above, for the second one the (NPH) and (NPT) collectives should be used. The (NPH) and (NPT) collectives would also be probably more adequate to describe the region near the surface, where the lattice of iron 
breaks its periodicity and might not be strong enough to counteract the internal pressure due to impurities. This work has been restricted to the (NVE) and (NVT) collectives, leaving the used of other collectives for future research.
% pt 4

\begin{figure}
\includegraphics[clip,width=0.99\columnwidth]{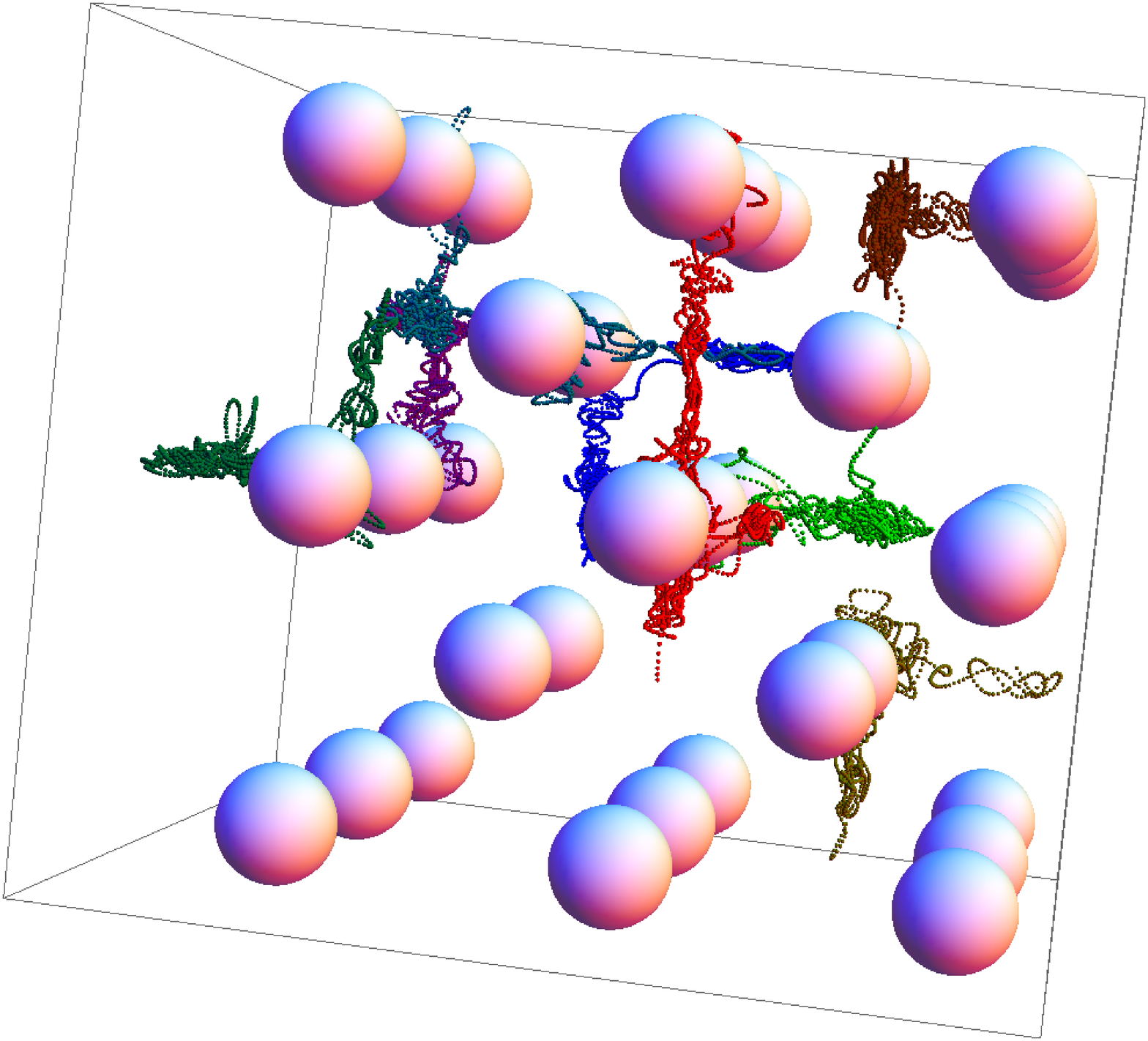}
\caption{(Color online) Simulated
random walks (T= 700 K) for eight 
interstitial hydrogen atoms simultaneously located in a 
$2 \times 2 \times 2$ bcc supercell. 
}
\label{Fig0}
\end{figure}

\begin{figure}
\includegraphics[clip,width=0.99\columnwidth]{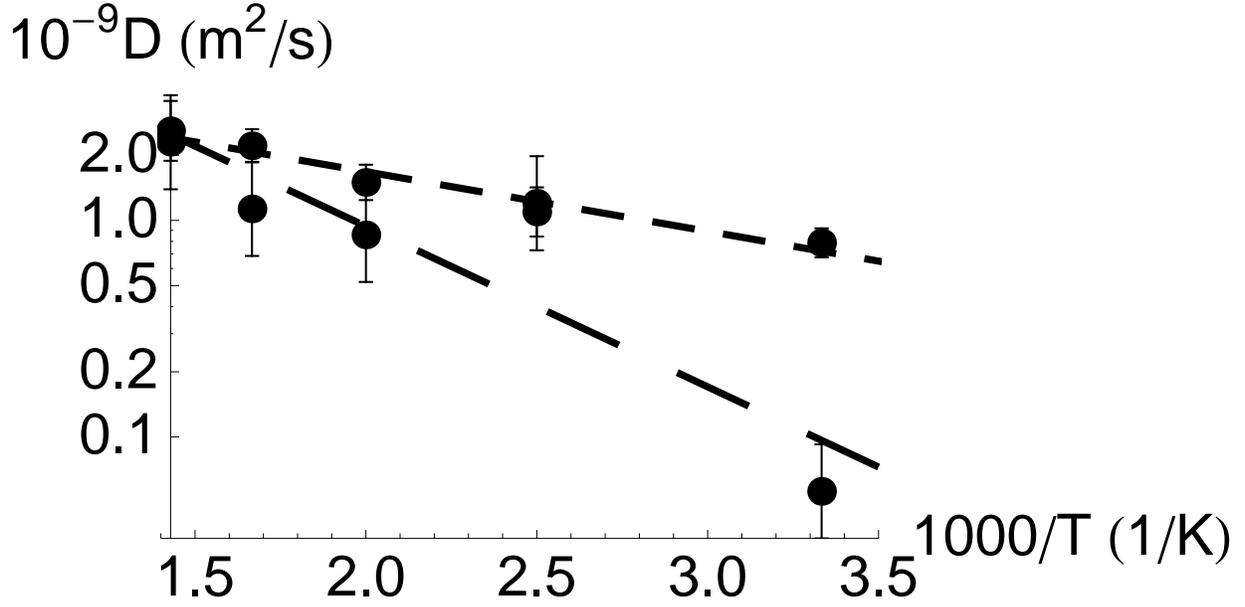}
\caption{
(a) (short dashed line)
Diffusion coefficient, D ($\frac{m ^{2}}{s}$), obtained for a
set of averaged diffusion MD trajectories for a single H atom in a 
$2 \times 2 \times 2$ bcc iron supercell.
A barrier for diffusion of $B_ {1}=0.145$ eV is obtained from a least-squares fit.
(b) (long dashed line)
Same for eight H interstitial atoms diffusing in the supercell.
A barrier of $B_{8}=0.047$ eV is deduced from the
fit.
Error bars have been estimated from standard errors 
($\frac{1.96 \sigma}{\sqrt{N}}$).
}
\label{FigArr1y8H}
\end{figure}

Diffusion coefficients are computed by assuming a random-walk for 
interstitial impurities.\cite{mehrer07} 
Fig.~\ref{Fig0} shows the different paths for the high density
case with eight hydrogen atoms diffusing simultaneously inside the
$2 \times 2 \times 2$ supercell. As already predicted by our static
geometry optimization interstitial hydrogen atoms avoid
each other and no tendency to formation of molecular hydrogen has
been observed.
We follow the different trajectories 
during the simulation time and compute averages
\begin{equation}
<|r(t)-r(0)|^2 > = 6 D t
\end{equation}
\noindent
to extract the three-dimensional diffusion coefficient D. 
Barriers for diffusion are estimated from an Arrhenius plot 
$
D = D_{0} e^{-B/k_{B}T}
$
by a least-squares fit.
Here, the prefactor $D_{0}$ is
related to a typical vibrational frequency for H in the 
Fe bcc lattice, and it is assumed to be a constant value
independent of the number of interstitial atoms in the
supercell.
This assumption is corroborated by our fits 
within an uncertainty of $\pm 12$\%
(Fig.~\ref{FigArr1y8H}).

\begin{figure}
\includegraphics[clip,width=0.99\columnwidth]{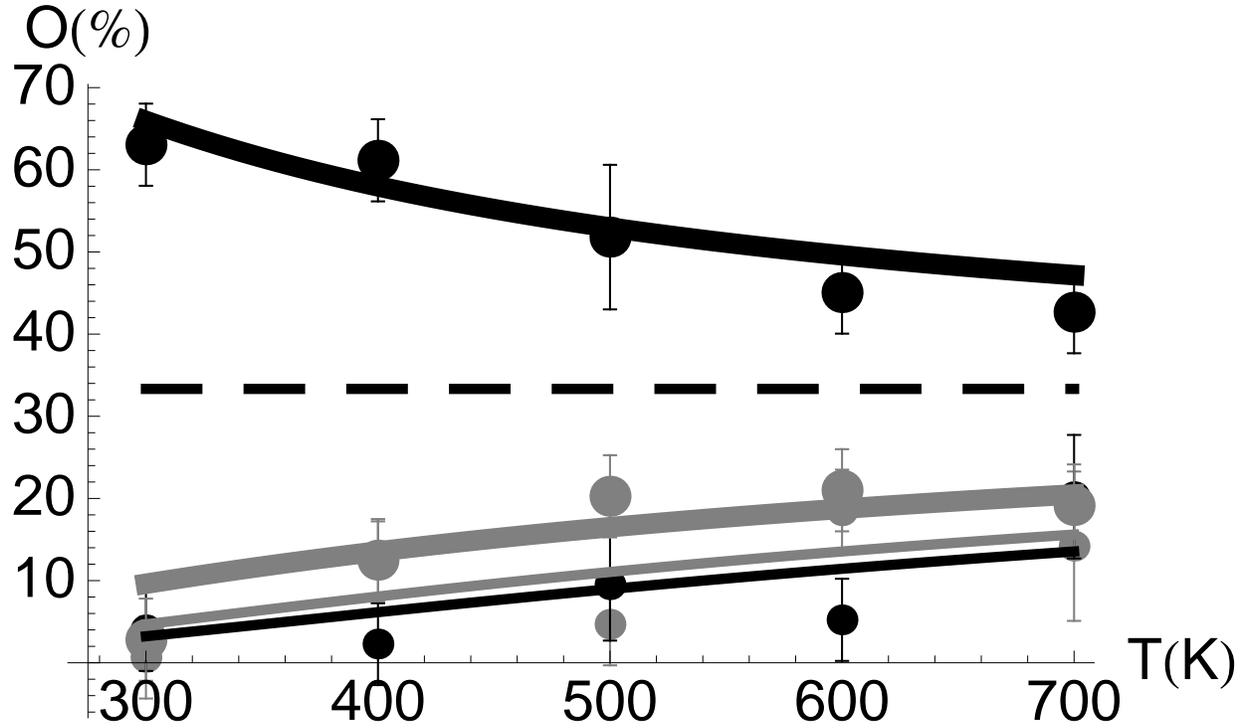}
\caption{
As a percentage over the total simulation time,
residence time around octahedral sites, $P_{O}$ for: 
(i) a single interstitial hydrogen 
(small gray dots;
thin gray line corresponds to $k_{B} \Delta=0.06$ eV in Eq.~\ref{EqSM});
(ii) two H (small black dots;
thin black line is for $k_{B} \Delta=0.07$ eV);
(iii) four H (large gray dots;
thick gray line is for $k_{B} \Delta=0.04$ eV); and
(iv) eight H (large black dots;
thick black line is for $k_{B} \Delta=-0.035$ eV).
The dashed line separates the two asymptotic
regions ($\Delta \ge 0$ and $\Delta \le 0$).
}
\label{Fig2SM}
\end{figure}

{\it Results}.
We simulate MD trajectories for a single H atom diffusing inside
the supercell to compare with previous results
derived from ab-initio DFT geometrical optimization and transition state 
theory.
From a least-square fit to the data in the Arrhenius plot
we determine a barrier for diffusion  $B_{1}=0.145$ eV 
(Fig.~\ref{FigArr1y8H})
in good agreement with previous values obtained under 
similar conditions (cubic symmetry, and the same interstitial density \cite{SanchezPRB08}).
This agreement shows that our MD simulations adequately sample the
relevant phase space. 
The time evolution of the interstitial atom can be further analyzed to show
that under this conditions trajectories near tetragonal sites (T) are
preferred over octahedral sites (O). 
 This conclusion can be made quantitative by computing 
a characteristic residence time for both sites. We assign parts of
the diffusing path either to T or O according to proximity to
estimate the likelihood to find the particle,
say, near O. 
Fig.~\ref{Fig2SM} displays the fractional occupation of octahedral sites
for 1, 2, 4 and 8 hydrogens in the $2 \times 2 \times 2$ supercell.
These values can be understood by comparing with a simple 
two-level model where the only parameter is the 
energy difference between T and O sites,  
$k_{B} \Delta=E_{O}-E_{T}$ :\cite{rosser}

\begin{equation}
\frac{O}{O+T}=\frac{1}{1+2e^{\Delta/T}} 
\label{EqSM}
\end{equation}

\noindent
This model reduces all the accesible volume for the interstitial hydrogen
diffusing in the unit cell to 
only a set of discrete lattice points (either T or O), but
in spite of its crudeness it already captures the essentials of the problem. 
Eq.~\ref{EqSM} has been plotted in Fig.~\ref{Fig2SM} for 
$k_{B} \Delta = 0.06, 0.07, 0.04$ and $-0.035$ eV, 
yielding least-square fits to the points extracted from the MD simulations
for $n_{H}=1, 2, 4$ and $8$ interstitial hydrogens, respectively. The 
dashed line represents the asymptotic equilibrium distribution 
($O_{T \to \infty} = \frac{1}{3}$)
to be approached from above or below depending on the
sign of $\Delta$. 
These results show that for $n_{H}=8$ ($\theta=\frac{1}{2}$)
$\Delta$ has moved from positive to negative and
the equilibrium site has been exchanged from T to O.
The dependence of
the parameter $\Delta$ with the interstitial density proves how
site adsorption energies
are affected by the presence of an increasing number of hydrogen atoms
concentrated in a particular region.   
This behavior follows the pattern previously predicted by ab-initio DFT geometry optimization
where the T site is the lowest energy configuration for low densities,
while for large densities occupancy of the O site favors a body centered tetragonal distortion
of the lattice and becomes the global minimum.
Although for the $\gamma$ phase, experiments reporting a qualitative
modification of the system around a hydrogen concentration of
$\approx 0.4$ show how the increasing density of interstitial impurities
might significantly modify the dynamics of these systems
(e.g., Fig. 13 in\cite{fukai05}). 
This is an observation that might help to
explain the spreading of  values extracted from different experimental  
techniques for diffusion barriers that has previously been
linked to partial occupation of both sites.\cite{kiuchi83} 
Partial occupation of both sites at a given temperature happens most naturally in molecular dynamics simulations, but it is not easy to describe in a standard geometry optimization.
We remark that the present approach represents 
a feasible route to investigate a regime that otherwise
is too difficult:
in the $2 \times 2 \times 2$ there are 48 different O-sites and 96 different T-sites, 
being the number of ways to distribute several interstitial among these combinatorially
large ($\approx 10^{11}$). Such a huge configurational space can only be addressed from
a statistical point of view and by letting the system to explore
as many relevant cases as possible by following its own dynamics.
To understand to which extent barriers for diffusion are affected by
the presence of extra interstitial
atoms we analyze the high-density case $\theta=\frac{1}{2}$
in more detail.
From an Arrehnius plot
(Fig.~\ref{FigArr1y8H})
we extract by a least-square fit 
an effective barrier of $B_{8}=0.047$ eV; 
significantly lower than the one found for a single 
interstitial, $B_{1}$. 

\begin{figure}
\includegraphics[clip,width=0.99\columnwidth]{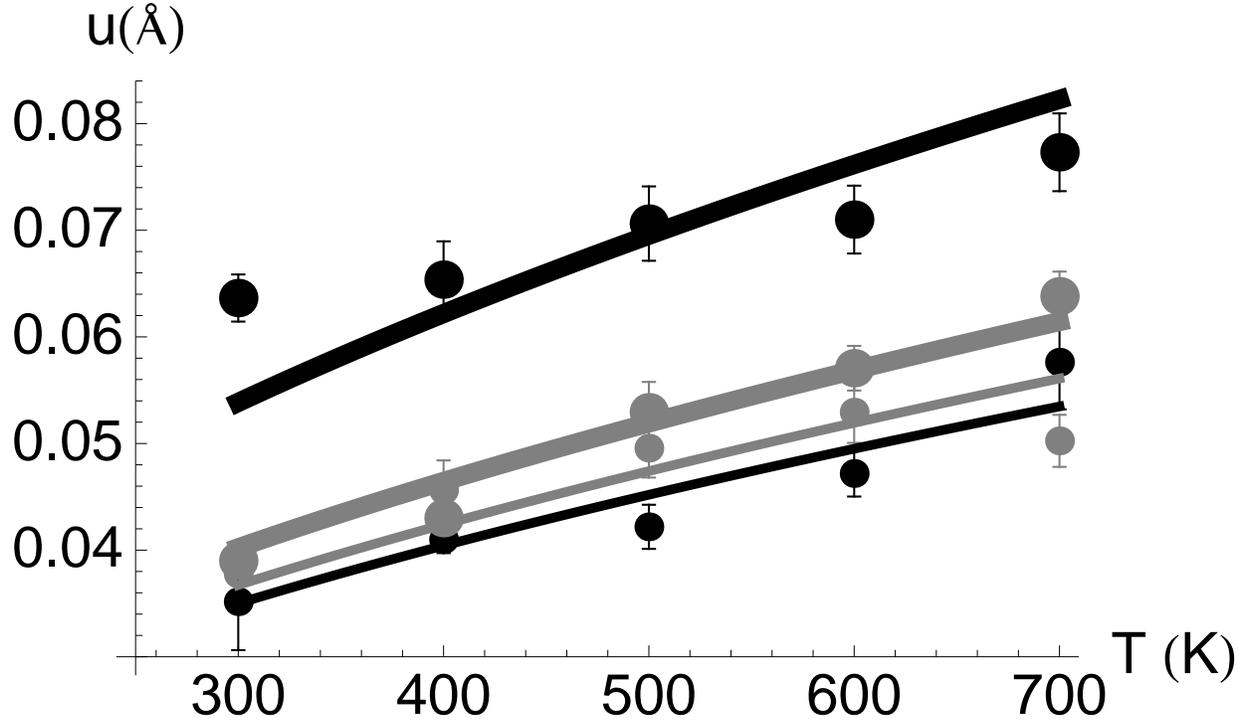}
\caption{
Iron root mean squared displacement from equilibrium positions 
u({\AA}) vs. temperature (K).
The density of interstitial hydrogen for each case is:
(i)  $n_{H}=1$ (small gray dots),
(ii) $n_{H}=2$ (small black dots), 
(iii) $n_{H}=4$ (large gray dots),
and (iv) $n_{H}=8$ (large black dots).
Corresponding continuos lines coded similarly in size and gray/black
are least-square fits to the
data using Eq.~\ref{EqDB} with Debye temperatures
$\Theta=381, 399, 360$ and $259$ K, respectively. 
}
\label{Fig3DBT}
\end{figure}

\begin{figure}
\includegraphics[clip,width=0.99\columnwidth]{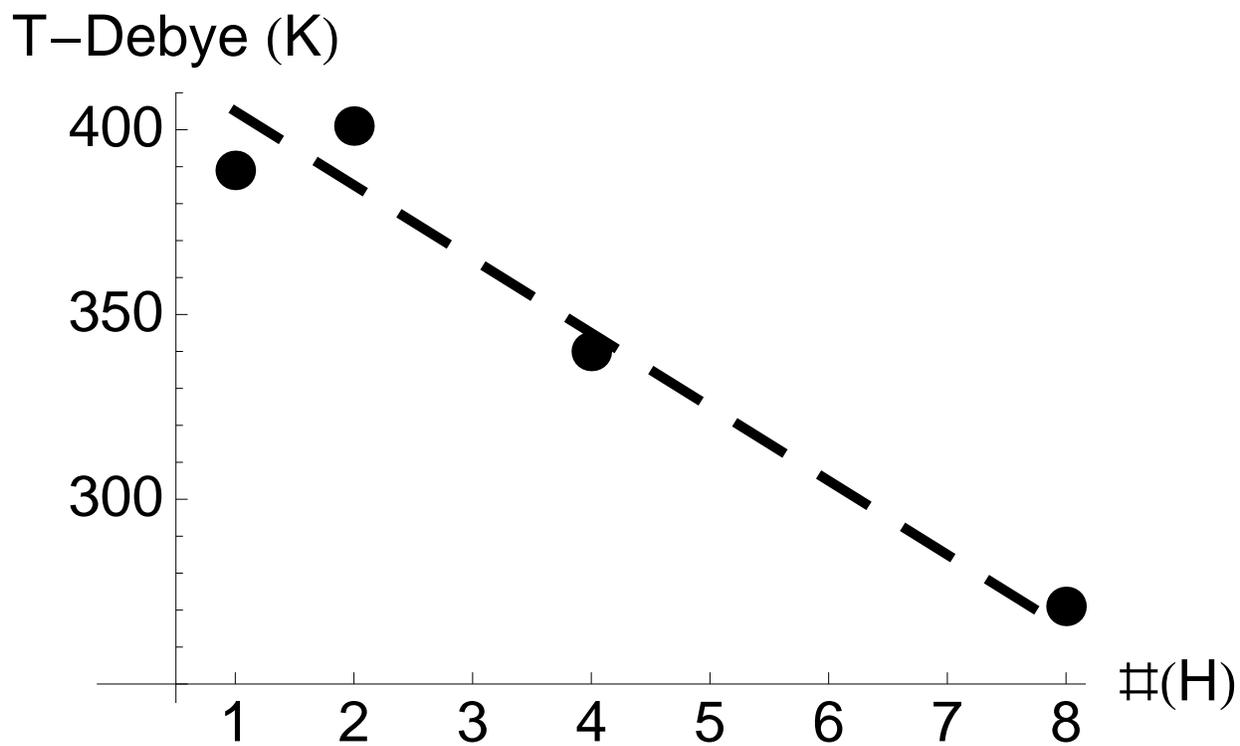}
\caption{
Debye-Waller temperature (K) 
vs. number of interstitial H diffusing in the unit cell. 
Dashed line is a linear least-squares fit to guide the eye.
}
\label{Fig4bisDBH}
\end{figure}

% referee point 1
Based in ab-initio DFT geometrical optimization we have previously suggested that an important consequence of interstitial hydrogen diffusing in $\alpha$-Fe is to weaken the Fe-Fe interaction.\cite{SanchezPRB08} From a physical point of view this effect is related to several factors: first, the presence of interstitial impurities increases the effective Fe-Fe distance and screens their interaction; second, the symmetry is distorted, an effect that we have found is important to explain the stability of octahedral sites in the high density regime; third, spin-polarized DFT calculations reveal that hydrogen contributes an extra spin to the system, but the total ferromagnetic moment does not grow accordingly. Current MD simulations should be taken as a highly controlled and clean theoretical experiment which results can extend these interpretations from $T=0$ K to a finite temperature, and from a limited amount of atoms sitting in the same supercell, to a larger number meeting together even if for a limited amount of time. It is interesting to observe that our MD simulations support a similar interpretation to the one derived from ab-initio DFT at $T=0$K.
% pt 1
We compute the mean square amplitudes of iron atoms vibrating
around their equilibrium positions, $<u^{2}>$, which is related
to the strength of the potential confining these atoms.
For each temperature we obtain the mean squared amplitude
of vibration
by fitting their time-averaged probabilities to a Gaussian distribution
centered around its equilibrium position. 
These values are compared with an isotropic Debye-Waller 
model for the mean squared displacement of atoms vibrating at
temperature T:\cite{james62}
% James pg 220 eq 5.69
% correcto, T y Theta en K, M en unidades de la masa del proton (o H).
%\begin{equation}
%<u>  \approx \frac{6}{\Theta} \sqrt{\frac{T}{M}}
%\label{EqDB}
%\end{equation}

\begin{equation}
<u^{2}>  = \frac{3 \hbar^{2}}{4 k_{B} M \Theta} 
(\frac{T}{\Theta}
 \int_{0}^{\frac{\Theta}{T}} \frac{x dx}{e^{x}-1}
+\frac{1}{4})
\label{EqDB}
\end{equation}

\noindent
where $\Theta$ is the Debye temperature and M
the mass of the atom.
Fig.~\ref{Fig3DBT}
shows how the 
root mean squared amplitude of vibration
($u$) increases steadily with the number of 
H atoms inside the $2 \times 2 \times 2$ supercell,
being nearly doubled from $n_{H}=1$ to $8$.
Adopting a Lindemann-like criterion
we can conclude that increasing the number of 
interstitial hydrogen drives the material
closer to a thermodynamic instability that eventually
should lead either to a phase transition or to the material failure.
This idea is more clearly illustrated by using
Eq.~\ref{EqDB} to fit these $<u^{2}>$ 
to an effective Debye temperature for each density
(Fig.~\ref{Fig4bisDBH}).
$\Theta$ decreases monotonously when the number of interstitial
hydrogens increases marking the softening of $\alpha$-iron, and
proving that the material is, at a given fixed T,
getting closer to its
own melting point under the internal pressure of dissolved H.
% referee point 2
This is closely related to the growing number of interstitial atoms sitting together in the same unit cell and the inverse situation (growing temperature getting closer to the melting point at a fixed number of interstitial impurities) cannot be inferred from our simulations because the Debye Temperature is constant under these conditions.
%2

{\it Conclusions.}
We have found by direct analysis of our MD trajectories
that the diffusing barrier for interstitial hydrogen inside $\alpha$-iron
depends on the density of diffusing atoms in the near region.
By using a simple statistical model we have also analyzed 
how the energy difference between the
T and O sites, $\Delta$, is modified by the presence of other interstitials.
Finally, by looking at the amplitude of vibration of iron atoms around
their equilibrium position, and comparing with
a simple Debye-Waller model, we conclude that 
the Fe-Fe interaction weakens as
the concentration of interstitial hydrogen increases,
finding that for the largest considered density the effective
Debye temperature for iron is already below room
temperature.

This work has been financed by the Spanish
CICYT (MAT2008-1497),
and MEC (CONSOLIDERS CSD2007-41 "NANOSELECT"
and "SEDUREC").

%\bibliography{fehmd} % Produces the bibliography via BibTeX.
% include *.bbl

\newpage


\end{document}